# Advancing Statistical Decision-Making in Sports Science

Janet Aisbett

*Meraglim Holdings Corporation, West Palm Beach, Florida, United States*

Eric J. Drinkwater

*Centre for Sport Research, School of Exercise and Nutrition Science, Deakin University, Geelong, Australia*

Kenneth L. Quarrie

*New Zealand Rugby, Wellington, New Zealand*

Stephen Woodcock

*School of Mathematical and Physical Sciences, University of Technology, Sydney, Australia*

**Summary.** The magnitude-based decisions (MBD) procedure was developed within sports science as an alternative to null hypothesis significance tests. It aimed to emphasise effect sizes and discourage dichotomous decision-making. The use of MBD was banned by some sports science journals following claims it lacks a theoretical foundation and leads to high Type I error rates. To address these claims, we first generalise contour-enhanced funnel plots to allow for ranges of meaningful effect sizes, then relate regions defined in these plots to the decisions made by MBD. We then mathematically show how MBD fits within a class of multiple decision procedures. We have implemented this theoretically sound version of MBD as a visualisation tool that supports generalised funnel plots. The use of MBD could encourage researchers to plan test directionalities, test levels and error definitions, and the visualisation tool may help stakeholders engage with the design of analyses and the interpretation of trial findings.

*Corresponding author:* Stephen Woodcock, School of Mathematical and Physical Sciences, University of Technology, P.O. Box 123 Broadway, NSW 2007 Australia.

**Email: Stephen.Woodcock@uts.edu.au.**

## 1. Introduction

A coach is considering a change to an athlete's weightlifting program, given exploratory findings that provide weak support for a small performance improvement but do not rule out a small performance loss. If an athlete is already likely to win a medal at a major competition, the coach won't risk the change. But if the athlete will probably finish just outside the medals, the coach might take the risk because a small improvement could push the athlete from a probably fifth to maybe third (a medal!) while a small deleterious effect would mean an inconsequential drop to seventh place or so. The coach's expert knowledge about the athlete and the competition must be brought together with the experimental results to make an informed decision. A conventional null hypothesis significance test at 0.05 alpha level and 80% power is not an appropriate decision aid for this coach.

While this and subsequent examples are set in sports science, the issues we highlight and the remedies we suggest have broader applicability. Our focus is on the applied scientist. Sports scientists working with teams and sporting organisations collect and analyse data on their athletes to improve athletic performance as well as to share their experience through scholarly publication. The interventions they study may be simple and cheap, such as changes to a pacing strategy, and are typically assessed through multiple measures of performance. Findings of small or uncertain benefit may still be of interest in a competitive environment, while results from previous studies may not transfer between specialised populations. Scientists works alongside coaching staff who are often poorly engaged in the research process but are responsible for the practical interpretation of findings. (Maughan et al. 2018; Bernards et al., 2017; Bishop et al., 2006; Lyle, 2018; Eisenmann, 2017).

A statistical procedure now known as magnitude-based decisions (MBD) was developed in this context. Batterham and Hopkins (2006: p. 50) describe it as providing "a more intuitive and practical approach" to estimating an effect size and to communicating uncertainty in results. MBD uses confidence intervals to emphasise imprecision rather than significance levels, and it identifies ranges of effects that are meaningful to the practitioner. The bounds of these ranges are applied in calculating *t*-statistics. To remove the focus from a dichotomous cut-off, *p*-values are reported both as numbers and as verbal labels that denote probability ranges. MBD (or MBI as it was formerly known) is implemented as Excel spreadsheets on a web domain sportsci.org which is dedicated to sports science. The procedure was adopted within that discipline, but not in other fields.

For over a decade critics have claimed MBD lacks a theoretical foundation, misuses Bayesian terminology, promotes small sample sizes and incurs high Type I error rates that are concealed from the user (Baker and Schofield, 2008; Welsh and Knight, 2015; Curran-Everett, 2018; Sainani, 2018; Sainani et al., 2019; Lohse et al. 2020). MBD was declared an unacceptable analysis method in a series of critiques and by some sports science journals and teaching institutions (e.g. Gladden, 2019; Harrison et al., 2020; Sainani et al., 2020). Statistical experts have been quoted in popular media saying "It's basically a math trick that bears no relationship to the real world" (Aschwanden, 2018) and "did not work" (Mannix, 2019).

Contrary to such claims, we will provide a sound theoretical basis for MBD within a class of multiple decision procedures that partition the parameter space through a family of hypotheses (Lehmann, 1957b; Finner and Strassburger, 2002; Goeman et al., 2010). Previous critiques of MBD largely followed Batterham and Hopkins (2006) in defining MBD labels through the placement of confidence intervals with respect to the smallest meaningful effect magnitudes (e.g. Baker and Schofield, 2008). The influential mathematical analysis by Sainani (2018) considers MBD as making binary decisions determined by constraints derived from the *p*-values for meaningful effect magnitudes; see also Lohse et al. (2020). Welsh & Knight (2014) use these *p*-values directly to interpret MBD as making four decisions. We will show these formalisations are incomplete.

We provide a visualisation tool for determining magnitude-based decisions, and it is this tool that we first describe. Visualisation has been long recognised as a mechanism for bridging the gap between scientists and stakeholders (content-experts), for example in exploring and interpreting clinical trial data (Gregson, 2015). Our tool is based on a generalisation of the funnel plots developed by Light and Pillimer (1984) to identify publication bias in meta-analysis. Funnel plots map trial summary statistics onto a chart in which typically the horizontal axis is effect size while the vertical axis is some measure of precision. Peters et al. (2008) introduced contour-enhanced funnel plots in which shading identifies regions of the chart that correspond to ranges of *p*-values for a standard test statistic. Distinguishing regions in this manner aids visual interpretation of the statistical significance of the findings (Sterne et al., 2017).

Apart from identifying asymmetry, which is taken as evidence of publication bias, funnel plot charts highlight studies that are significantly different from the pooled mean and identify "small-study effects" when these tend to differ from findings from larger studies. Funnel plots are also used to understand variability of a performance indicator amongst organisations or localities, and to describe sample size effects in statistics education (Mazzucco et al., 2017; Thomas et al., 2012). Criticisms of funnel plots include changes to their appearance (and hence visual interpretation) with different precision measures, and spurious asymmetry when data transformations produce interactions between effect sizes and standard errors (Terrin et al., 2005; Lau et al., 2006). In meta-analysis, they are therefore recommended as an adjunct to other tools.

A deficiency of funnel plots that to our knowledge has not previously been considered is their lack of support for distinguishing effects that have practical meaning from those that are statistically significant but not meaningful. While the desirability of testing for meaningful effects has been promoted for over 60 years (e.g. Hodges and Lehmann, 1954), moving researchers beyond standard null hypothesis significance tests (NHSTs) for zero effects has proven difficult. Our first task is therefore to generalise (contour-enhanced) funnel plot charts to accommodate the notion of ranges of effect sizes that are meaningful to practitioners.

Our paper is organised as follows. Section 2 presents our generalisation of contour-enhanced funnel plots and suggests how this might aid the design of analyses and the interpretation of trial findings. Section 3 relates the funnel plots to the magnitude-based decision procedure and examines MBD's controversial error rates and terminology. Section 4 gives a theoretical foundation for MBD as a decision tool by first reviewing Lehmann's (1957a,b) theory for a class of multiple decision procedures, then presenting a novel construction to accommodate MBD's testing of hypotheses at multiple levels. Section 5 concludes that these decision tools are helpful for applied scientists and that previous critiques of MBD have taken too narrow a perspective of its potential usage.

## 2. Generalized Enhanced Funnel Plot

### 2.1. Generalising the funnel plots

A conventional funnel plot divides the chart of effect sizes versus standard error into three regions, corresponding to inconclusive findings and those that are significant with size greater than or less than the pooled mean (or zero, if centred symmetrically). Fig. 1a depicts a typical funnel plot, displaying findings from a meta-analysis by Grgic et al. (2018: Table 2). The horizontal axis represents Hedge's *g* derived from measures of maximal strength performance when athletes ingested caffeine versus placebo. The numbers 1 to 6 correspond to the alphabetical order of the first six studies reviewed. Here and in all subsequent figures, the range of non-meaningful effects is assumed to be symmetric about zero and significance levels are with respect to one-sided *t*-tests with 18 degrees of freedom; the overall pictures change little with degrees of freedom down to 10 or so, or for larger values.

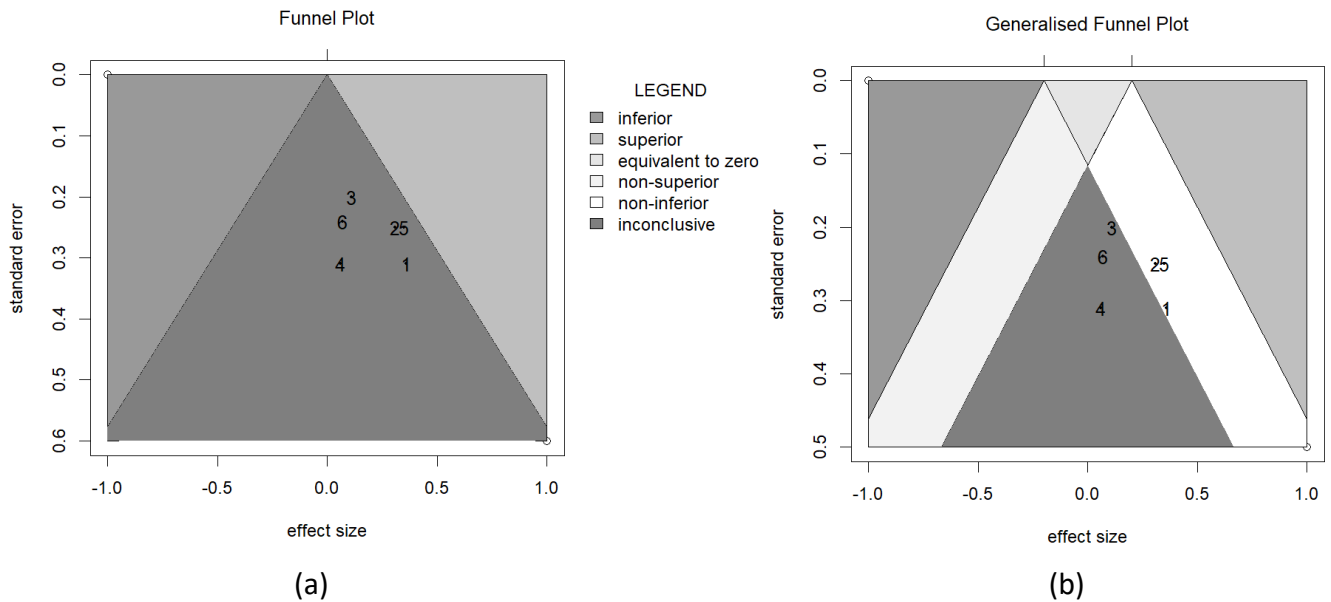

(a)                                                            (b)

**Fig. 1.** (a) Standard funnel plot. (b) Generalised funnel plot when the range of effects equivalent to zero is symmetric about zero.

To interpret Fig. 1b, consider the superposition of two funnel plots centred on each of the bounds of meaningful effect sizes (shown here with magnitude 0.2, i.e., small on Cohen's scale). Suppose each of the funnel plots uses the same test level, here set to a one-sided alpha level of 0.05. The region labelled 'inconclusive' in Fig. 1b is the overlap of the 'inconclusive' regions of the two plots where the tests are not significant. The region labelled 'equivalent to zero' is the overlap of the 'superior' region of the plot centred at -0.2 and the 'inferior region' of the plot centred at 0.2, and hence is where the effect is conclusively in the range [-0.2, 0.2] of effect sizes that are not meaningful. The region labelled 'superior' in Fig. 1b is where the superior regions overlap, and similarly for 'inferior'. 'Non-inferior' and 'non-superior' are where one, but not both, plots are inconclusive. This terminology is that used in comparative drug trials (e.g. Aberegg et al., 2018) because the regions in the chart correspond to the rejection regions of the tests used to establish superiority and so on.

Fig. 1a shows none of the study findings from Grgic et al. (2018) are significantly different to zero, so none are significant in tests for meaningful effect sizes. However, we see from Fig. 1b that studies 1, 2 and 5 find the treatment is non-inferior to placebo.

The six regions shown in Fig. 1b can also be interpreted in terms of the positions of confidence intervals (CIs) (Baker and Schofield, 2008). The inconclusive region contains all findings whose CIs cross both boundaries of the range equivalent to zero; the non-inferior (resp. non-superior) region is where the CI crosses the upper (resp. lower) boundary of this range; the equivalence region is where the CI fits within the range; and the superior (resp. inferior) region is where the CI only contains positive (resp. negative) meaningful effects.

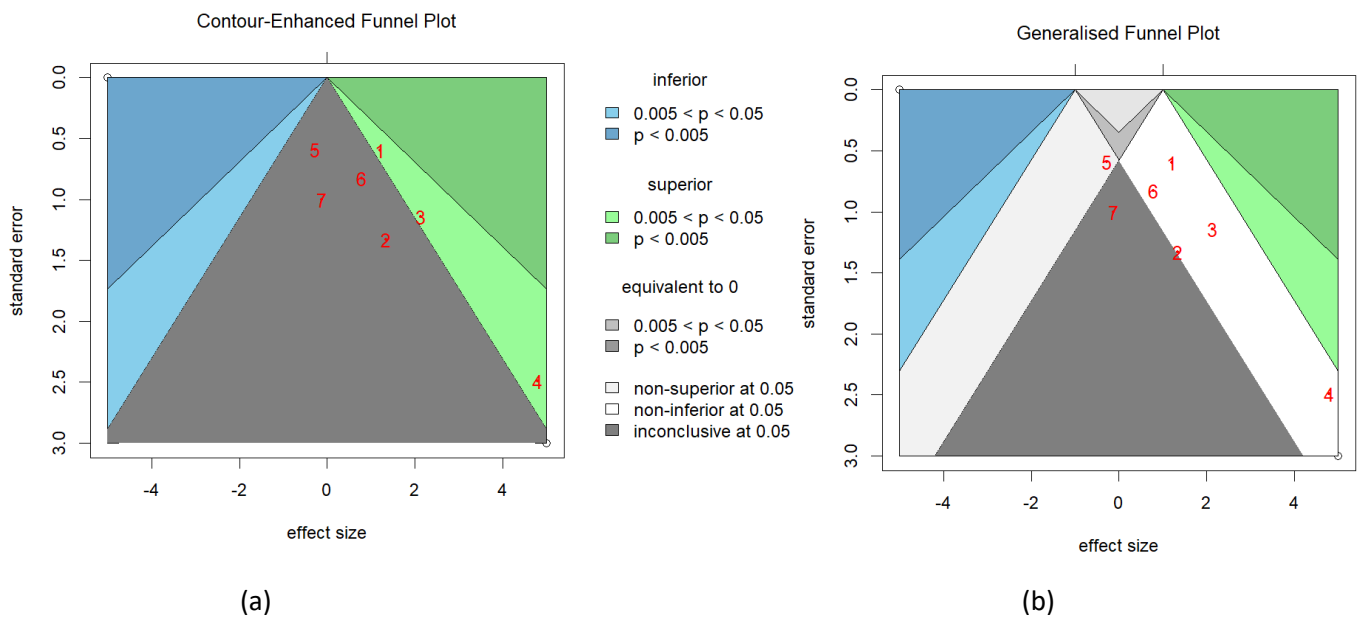

(a)                                                                 (b)

**Fig. 2**. (a) Enhanced funnel plot. (b) Generalised enhanced funnel plot.

Contour-enhanced funnel plots can similarly be generalised to highlight meaningful effect sizes. Contour-enhanced plots typically consider two *p*-values, giving three ranges: say, the *p*-values from one-sided NHSTs that are greater than 0.05, less than 0.005, or in-between. Fig. 2a illustrates, using these ranges. The charts are centred on zero, and they contain five partitions of the chart area, corresponding to effects that are not significant at the least stringent test level plus two regions each for positive effect sizes and negative effect sizes. Fig. 2b shows a contour-enhanced plot adapted to highlight meaningful effects, again assuming the smallest meaningful magnitudes are symmetric about zero. There are now nine regions. The numbered plots in this figure are findings from a single intervention reported in Bartolomei et al. (2014) which we discuss in Section 3.

## 2.2.    Using the generalised funnel plots for design and reporting of studies

The generalised funnel plots can be employed at the design stage as well as in interpretation of study findings, using *a priori* estimates of anticipated effect sizes and variances, as required in a conventional Neyman-Pearson analysis, together with planned sample sizes. The funnel plots therefore can assist researchers move beyond the blind application of NHST, as we now discuss with suggestions as to how stakeholders (coaching staff and team officials in the sports science context) should be involved.

### 2.2.1 Taking account of meaningful effect sizes

Thinking about which effect sizes are practically meaningful is the first step to moving beyond NHST. Documentation at the sportsci.org site explains how to select the smallest meaningful effect sizes in specific sports science scenarios, e.g. to gain Olympic gold (Hopkins 2010). However, this selection depends on the level of athletic participation, with a small benefit potentially important in a competitive environment (Maughan et al., 2018). Thus, in strength training Rhea (2004) finds standardised effect sizes less than 0.5 are not meaningful for untrained subjects, but this threshold lowers to 0.35 for recreationally trained athletes and to 0.25 for highly trained athletes. Unless the populations are comparable, existing literature may therefore be of limited use in estimating smallest meaningful effect sizes. Indeed, the meta-analysis by Lohse et al. (2020) of studies using MBD analyses found most researchers applied default settings.

Lakens et al. (2018) emphasise the need to justify choices of smallest meaningful effect sizes rather than use defaults such as a Cohen's d of 0.2 and Cohen (1988) himself advises that defaults for considering effects to be "moderate" and so on should only be used when there is no better justification. In the context of sports science, Buchheit (2016) stresses the importance of meaningful effects to stakeholders. To avoid resorting to defaults, this suggests researchers use expert specialist knowledge. While no one stakeholder may be able to determine the appropriate threshold, a group of them will do better than the statistician alone in answering questions such as "Would you consider a one degree increase in range of motion of your knee to be a meaningful sign of recovery?" A large body of literature exists on methods to elicit such information, including the well-known Delphi method (Dalkey and Helmer, 1962) and discrete choice experiments (Terris-Prestholt et al., 2019).

### 2.2.2. *Staying open-minded about the direction of tests and the definition of errors.*

If researchers assume positive effects are all that are of interest they may overlook findings that are useful to the stakeholder. The generalised funnel plots charts show results from tests for inferiority, superiority, non-inferiority, non-superiority and equivalence to zero. These tests can be conducted simultaneously without adjusting for multiple effects (Berger, 1982; Goeman et al. 2010).

For example, suppose a researcher hopes to establish superiority but anticipates a small effect size, say twice the smallest meaningful. Fig. 2b shows that the anticipated standard error needs to be about half the smallest meaningful effect size to plausibly seek a significant result at level 0.05. Establishing non-inferiority may be enough for a stakeholder's purpose and will be a more realistic aim if sample sizes are constrained.

Conventional Type I and Type II error rates are only defined in relation to a single hypothesis of interest. Neyman (1977) recommends that this hypothesis be set so that the errors deemed more costly by the stakeholder are Type I, because stronger control is exercised over these than over Type II errors. For instance, if it is deemed more important to not miss out on faster recovery than it is to spend 10 minutes on a new stretching routine that brings no additional benefits, then the hypothesis of interest should be flipped from superiority (meaningful benefit) to non-superiority. The test that must be rejected is then that the routine is beneficial, and a Type I error is incurred when a beneficial effect is rejected. In this situation, the intervention will be implemented (or will continue to be used) *unless* there is evidence that it does not work.

Elicitation techniques allow stakeholders' preferences to determine the direction of a test, and from that define Type I error and Type II errors. To reiterate, when the stakeholders' perceived loss from falsely rejecting a beneficial effect exceeds that from falsely failing to reject a non-beneficial effect, Type I errors should be made from deciding there is no benefit when there is benefit, rather than vice-versa. Indeed, a coach might not attach *any* loss, and hence not attach any error, to keeping up a stretching routine, say, when in fact it did not aid recovery, because of other perceived benefits. A decision to stop the practice would plausibly require finding no benefit across all measures of interest, or else finding that harm might be caused on some measure.

### 2.2.3. *Moving beyond p<.05*

Researchers have long been encouraged to move beyond *p*<.05, with little impact (Wasserstein et al., 2019). By displaying rejection regions for tests at two or more alpha levels, contour-enhanced funnel plots force researchers to think about test levels. Johnson (2019) recommends moving from 0.05 to stricter levels like 0.005; on the other hand, to compensate for expected limited sample sizes the weak alpha level of 0.25 may be recommended, as for studies into the shelf-life of drugs (ICH, 2003).

Take the example from 2.2.2 but look now at Fig. 3a, which includes rejection regions at level 0.25. We see that superiority might realistically be established at this level if the anticipated standard error is four-thirds of the smallest meaningful effect size. Gathering the required sample size (dependent of course on the research design and anticipated variance) may now be feasible. Presenting the findings as very weak evidence of superiority alongside moderate evidence of non-inferiority may provide useful additional information to the stakeholder.

Always, the acceptability of error rates depends on context (Betensky, 2019). When sample sizes are limited, consideration needs to be given to the balance with false negatives (Mudge et al., 2010). Type II error rates are often overlooked even though they may be important to the stakeholder. For example, Lohse et al. (2020) make no mention of Type II error rates when reworking results from a study using an MBD analysis that drew conclusions from tests at alpha level 0.25. If balance of error is considered, and typical assumptions are made about variance and so on, this weak level can be shown to be preferred over the standard level 0.05 adopted in the rework.

### 3. MBD in Relation to Generalised Funnel Plots

Armed with generalised funnel plots, we now provide a visually based description of MBD. The procedure is currently implemented on the site sportsci.org in a series of spreadsheets, supported by documentation. These spreadsheets cater for different trial designs, e.g. post-only crossover. Because there has been no full formal explanation of the procedure, we use as definitive source the computations in the MBD spreadsheets (excluding sportsci.org/resource/stats/xcl_Bayesian.xls). We then give an example of MBD usage, before discussing error rates and terminology that have been central issues in criticisms of the procedure. Section 4 will formally describe the spreadsheets' computations and spell out minor changes needed to make them conform to the theory and to our chart-based description.

### 3.1. MBD as a generalised funnel plot

The decision returned by the MBD spreadsheets essentially corresponds to the region in which a plotted point lies in a generalised enhanced funnel plot when three rather than two $p$-values are used. MBD uses the default values of 0.25, 0.05 and 0.005, representing a spread from unusually weak to strong test levels (Hopkins, 2007). Fig. 3a illustrates such a generalised funnel plot.

A generalised funnel plot, however, does not completely correspond to MBD because MBD imposes a constraint on its decisions: in the terminology we have been using, the decision is called inconclusive unless either non-inferiority or non-superiority can be established at a specified level. The specified levels are designed to suit specific sports science contexts, giving two versions of MBD that are (somewhat unhelpfully) named non-clinical and clinical MBD.

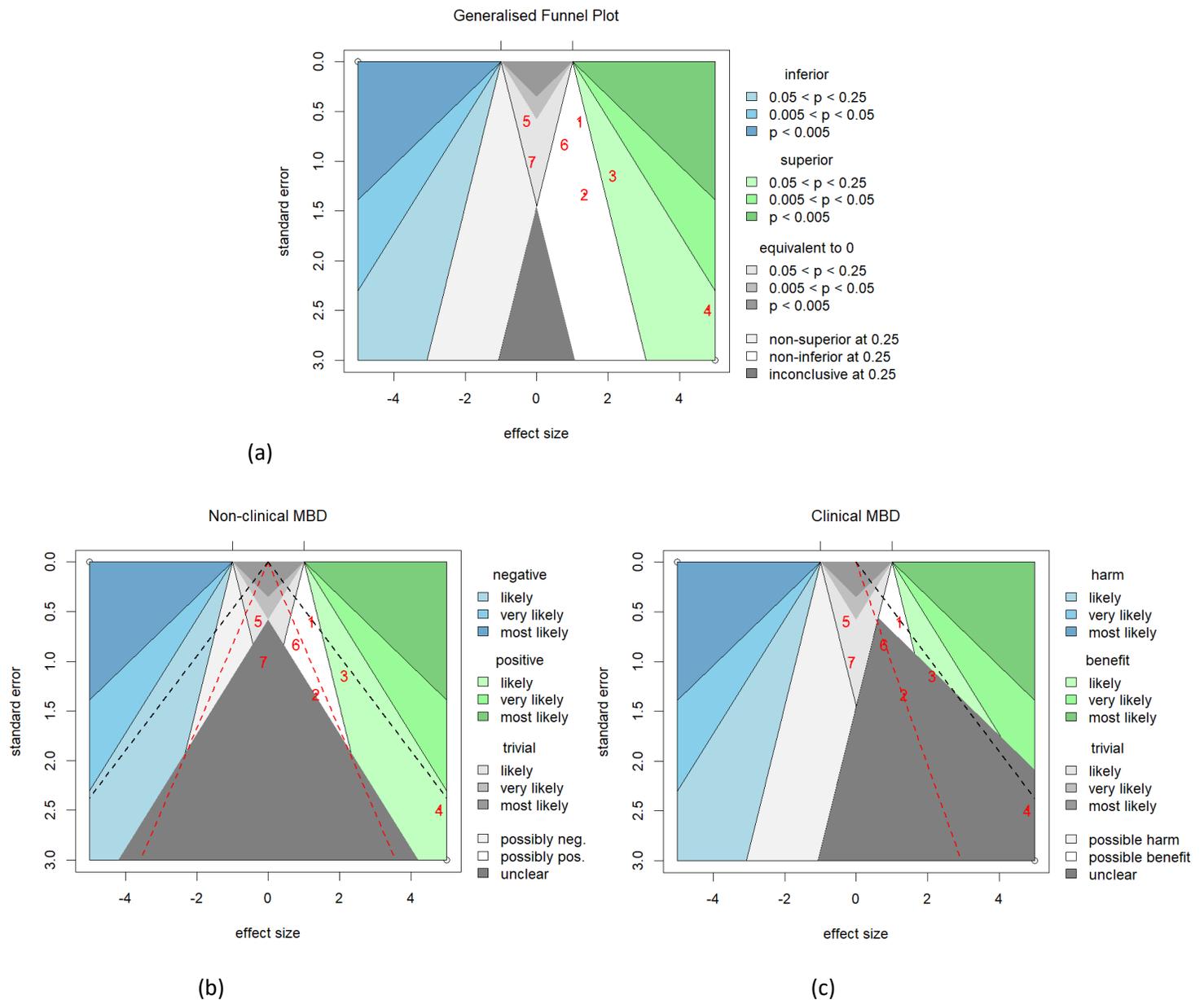

(a)

(b)                                    (c)

**Fig. 3**. (a) Generalised funnel plot with *p*-values at 3 levels; (b) and (c): Decision regions and labels used in non-clinical and clinical MBD.

Fig. 3b illustrates non-clinical MBD. Decisions must satisfy either a test of non-inferiority or of non-superiority at an alpha level of 0.05. Thus if, say, superiority is decided at the unusually weak one-sided test level of 0.25, the decision still guarantees that non-inferiority has been established at a stronger level. The conclusion that the effect numbered 7 is trivial at the very weak level of 0.25 is thus overruled to be unclear (inconclusive) in MBD (Fig. 3b).

Non-clinical MBD is typically applied to comparative investigations of two treatments on a population or a treatment on two populations, where improvement on measures of interest is anticipated in both. In contrast, clinical MBD (Fig. 3c) assumes that effects smaller than the negative bound on meaningful effects are harmful and effects larger than the positive bound are beneficial; that avoiding harm is more important than missing benefit; and that not missing benefit is more important than wrongly assuming benefit. As a result, conclusive decisions of benefit require inferiority (harm) to be rejected at the strong default level of 0.005. If this is satisfied, then MBD spreadsheets return a decision 'consider using' for findings of possible benefit (which is merely non-inferiority when no other decision can be established at any level) as well as for findings of benefit at level 0.25 and higher.

If the assumptions behind clinical MBD were appropriate for the intervention whose findings are depicted in Fig. 3, it would not be considered safe to recommend the intervention on the basis of effects 3 and 4 because a deleterious effect on these measure has not been ruled out. For the coach described in our Introduction, prepared to take a greater risk of performance loss, the harm test alpha level of 0.005 would need to be relaxed.

The key thing to take away from Fig. 3 is that the three charts differ only because the region representing inconclusive findings in 3a is enlarged under MBD. The language in the legends has also changed. The definitions of the MBD labels are approximately determined by comparing the legends of Figs. 3b and 3c with that of Fig. 3a. Essentially, the MBD terms 'most likely', 'very likely' and 'likely' correspond to test alpha levels 0.005, 0.05 and 0.25 respectively. The translations are made more complex by the notion of 'unclear' findings, which affect 'likely' decisions (and 'very likely beneficial' decision in the clinical case) with larger standard errors. Section 4 and the Web-Based Supporting Material provide precise definitions of the 12 decisions represented by the 12 regions identified in Figs. 3b and 3c.

The dashed lines in Figs. 3b and 3c are the boundaries of NHST rejection regions. The black lines are those for one-sided tests at level 0.025. The red dashed lines are described below.

### 3.2.    Example of MBD use in sports science

Bartolomei et al. (2014) is typical of sports science studies to which MBD analyses have been applied. The 24 participants were male athletes competing at high level in sports involving extensive resistance training (e.g., shot put, American football). They were divided into two groups that performed the same suite of training exercises over the same session lengths for 15 weeks; however, one group had their training structured to increase intensity in 'blocks' every few weeks, while the second group had a more traditional training in which intensity cycled in waves over the 15 weeks. Pre and post measurements were recorded of seven strength and power tests. Bartolomei et al. (2014; Table 2) set the magnitude of the smallest meaningful effects to 0.2 times the standard deviation of the pre-trial measurements over all participants. Because this varied with the measure, we normalised the summary statistics (standardised mean difference, standard error) by the relevant smallest meaningful effect before plotting the findings in Fig. 3.

The axes are therefore in units of the smallest meaningful effect magnitude, so the bounds of the meaningful effect ranges are -1 and +1. For each of the measures, the positive direction represents a relative improvement for the block protocol.

Bartolomei et al. (2014) applied non-clinical MBD since the trial was comparing two interventions which both brought overall improvement on the measured indicators. Fig. 3b shows all findings are under the dashed black lines representing the NHST boundaries in Fig. 3b, so none are significant at a one-sided alpha level of 0.025. The data points 1 to 4 represent findings from upper body measures, and the figure shows that the block protocol is non-inferior for these at level 0.05, and for one of the other measures. The block protocol is relatively superior for two of the upper body measures at alpha level 0.25 whereas even such weak evidence of superiority could not be concluded for the traditional protocol for any measure. Despite the lack of significant results under standard NHST, a coach might decide to use the block protocol given the MBD findings of weak evidence for benefit and low chance of harm, since there is little cost associated with switching the training strategies.

### 3.3.    MBD controversies

Here we investigate claims of high Type I error rates and whether the MBD probability-related labels are prone to misleading users to believe these rates are lower than they are.

#### 3.3.1. Are MBD Type I error rates unacceptably high?

To assess the claims of high MBD error rates, we use the bird's-eye view provided by the visualisation charts, removing the need for case-specific simulations or numerical approximations. Put simply: if the rejection region of test A contains the rejection region of test B, test A must have higher Type 1 errors regardless of the true effect size or the range of meaningful effect sizes.

In claiming MBD produces unacceptably high Type I error rates, critics have generally assumed that the hypothesis of interest is superiority and thus call a Type I error whenever the nominated test rejects an effect that is not meaningfully positive (Sainani, 2018; Sainani et al., 2019, Lohse et al., 2020). As we said, Fig. 3 assumes the unit on both axes is the smallest meaningful effect magnitude. Fig. 3b shows that the NHST rejection region delineated by the black dashed lines includes the non-clinical MBD rejection region of the test for superiority at alpha 0.25 for standard errors up to about 0.7 times this magnitude. This multiplier rises to 2.5 or more with an alpha of 0.05. Below these crossover points, NHST will be expected to have higher Type I error rates than MBD; above these crossover points, MBD has higher expected Type I error rate. With group sizes at the median of 10 found in sports science studies reviewed by Lohse et al. (2020), larger standard errors would be expected so MBD with test level 0.25 would have a higher Type I error rate.

Clinical MBD documentation describes 'consider using' as meaning an intervention can be considered not harmful but is not necessarily beneficial (Hopkins, 2020). Nevertheless, critics have read this as superiority being the hypothesis of interest. If we do the same, and compare the rejection region for a one-sided NHST at level 0.025, bounded by the dashed line in Fig. 3c, we see that the Type I error rate is expected to be higher for NHST at larger standard errors (smaller sample sizes) as well as small standard errors (large sample sizes). In-between, the MBD error rate is higher.

The lower error rates for clinical MBD at the standard errors associated with small sample sizes common in sports science were not noted by Sainani (2018). Instead, when discussing clinical MBD, critics have focussed on the sample sizes recommended by Hopkins (2006). These sample sizes can be shown to give an expected scaled standard error at the apex of the regions labelled 'unclear' in Figs. 3b and 3c (ignoring slight differences in the diagrams with the higher degrees of freedom). For non-clinical MBD the expected Type I error rate will be *lower* than for standard NHST. Claims about high Type I error rates at these recommended sample sizes are specific to clinical MBD, a distinction that has not been made clear.

Generalisations about MBD incurring or allowing higher error rates than standard NHST are not possible. The red dashed lines through the origin in Fig. 3 are the boundaries of one-sided NHST rejection regions with alpha level 0.125 (Fig. 3b) and 0.17 (Fig. 3c). These boundaries are entirely outside the rejection regions corresponding to the decisions 'likely positive' (Fig. 3b) and 'consider using' (Fig. 3c) and so the associated Type I errors must always be lower. We conclude that for a true zero effect the non-clinical MBD Type I error rate when treating 'likely positive' as substantive will never exceed 12.5%. The clinical MBD error rate given a true effect of zero will never exceed 17% even when 'possibly beneficial' findings are called beneficial. These bounds apply for all degrees of freedom. The non-clinical rate must be doubled if the hypothesis concerns all meaningful effect sizes.

These are unusually high Type I error rates but whether they are "unacceptably high" (Sainani, 2018: p2166) is entirely dependent on context. If they are, then the decision criterion should be changed. When benefit in clinical MBD is concluded at 'likely' level, the Type 1 error rate for a true zero effect is capped at 5%. When the criterion for concluding a positive or negative result in non-clinical MBD is 'very likely' (as recommended in MBD documentation; Hopkins, 2020), this rate doesn't exceed 5% for standard errors less than three times the smallest meaningful effect magnitude; with 'most likely' this rate never exceeds 0.5% for standard errors less than 20 times this magnitude.

Sainani (2018; p2174) claimed the dependence of MBD Type 1 error rates on sample size and on the smallest meaningful effect was a flaw that hid error rates from the user. However, only the specific assumptions used in standard NHSTs allow Type I error rates to be equated with alpha levels; whenever the nominated true effect differs from the bound of a one-sided hypothesis, Type I error rates under a *t*-test depend on the standard error. We have presented upper bounds on these error rates for MBD when the true effect is zero. The tool that confirmed these, and which created Fig. 3, can be used to further explore Type I error rates with different nulls (ja090.shinyapps.io/MBDSS).

### 3.3.2. Are MBD labels misleading?

After an MBD decision is made algorithmically, its suitability must be assessed by stakeholders. So in considering MBD in relation to conventional statistical analyses, the role of stakeholders as well as researchers need to be considered and "results need to be articulated to the key stakeholders at an appropriate level of detail/language so that all understand" (Bishop et al., 2006: p. 163). MBD describes findings informally in terms of the decision about the effect size and the level of evidence attached to that decision though the $p$-value. Sainani and colleagues claim that these MBD descriptors are not transparent and lead users to over-optimistic conclusions (Sainani, 2018; Sainani et al., 2019, Lohse et al, 2020; Sainiani et al., 2020).

According to frequentist interpretation, the event to which the MBD level of evidence is attached is the occurrence of the observation given a true effect *not* in the postulated size range. The MBD 'likely positive' label is thus attached to a decision when data at least as extreme as that observed could occur with probability in the range 5-25%, given a true effect that is not meaningfully positive.

While experiments show people differ in their quantitative interpretations of terms such as 'very likely' and 'unlikely' when these are used to represent the rate of occurrence of an unspecified event (Lichtenstein and Newman, 1967), there is a broad agreement. For example, Hillson (2005) reports 'highly unlikely' was on average experimentally mapped to the range 0-15%, 'unlikely' to 15-45%, 'possible' to 29-58%, 'likely' to 50-68% and 'highly probable' to 64-78%, while Conrow (2003) gives the corresponding ranges as 5-15%, 15-25%, 35-65%, 65-85% and 75-85%. This suggests that the 'likely' MBD label is applied to observations that people would judge 'unlikely' or 'highly unlikely' to occur when the effect is outside the postulated range.

The inversion fallacy (Villejoubert and Mandel, 2002) leads people to interpret a finding that an error is 'unlikely' when the true effect is outside the given range as meaning that the true effect is 'likely' to be in it, and vice-versa. This transformation may be due to neglect of base rates (Bar-Hillel, 1980). In any case, when told that the trial outcome is 'likely positive' and then asked the rate at which an effect outside the postulated range would be wrongly found to be positive, most people would presumably nominate a figure no more than 25%. Conversely, if told the long-term rate of mistaking a true value outside the postulated range was 25%, then most people would conclude that the true effect was 'likely' in it. While to our knowledge no-one has conducted experiments in relation to this claim, the literature supports the contention that the MBD label does not misrepresent the true error rate. Rather, the typical interpretation of the label and the error rate would generally be consistent because of the inversion fallacy.

This amalgam of Bayesian and frequentist thinking could be avoided by a Bayesian re-analysis. MBD documentation generally describes the procedure as semi-Bayesian or reference-Bayesian, and this has been used to justify the probabilistic terms. However, establishing a realistic informative prior may be difficult in the applied sports science setting described in the introduction. Alternatively, results could be presented using common language terminology designed to avoid the base rate or inversion fallacies. For example, an MBD 'likely positive' finding would be presented as 'unlikely if the true effect is not positive'. The double negatives invite readers to informally invert them, and so this terminology may not be worth the attempt at correctness. Another alternative is to use non-probabilistic terms such as 'weak', 'moderate' and 'strong' which are used in the MBD spreadsheet sportsci.org/resource/stats/xcl.xls (Hopkins, 2020). However, it is not known how such terms would be interpreted.

Overall, the MBD terminology appears useful. Unless sports scientists have good communication with coaches, says Buchheit (2017: p. 40), "their fancy reports with high quality stats will end up in the bin" and he illustrates how MBD's terminology can help. Of course, the researcher must understand the correct interpretation and use that correct interpretation in scholarly reporting.

We believe the label 'possibly positive' for the decision when negative is rejected at a level of 5% (non-clinical) or 0.5% (clinical) is arguably open to misinterpretation. Hopkins and Batterham (2016) correct this by using the label 'positive-trivial'; and likewise 'negative-trivial'. The need for this clarification is supported by a review of 232 studies that applied the MBD procedures (Lohse et al., 2020). This review claimed 63 studies using the non-clinical variant treated a 'possibly positive' as positive. While this count should be treated with great caution—it includes findings in which the conclusions were aggregated (as in Bartolomei et al., 2014) or were qualified or for which NHST or other tests were the basis for making conclusions—it seems advisable to spell out these parameter ranges as 'positive or trivial' and 'negative or trivial'.

## 4. MBD as a Multiple Decision Problem

Here we describe MBD in terms of a theory of the multiple decision problem (Lehmann, 1957a,b). This provides a formal foundation for the procedure, subject to a change of relatively minor practical consequence. Our notation varies from that of Lehmann, and as before our formal description reflects the MBD spreadsheet computations rather than documentation produced by the MBD developers.

### 4.1. Lehmann's theory of multiple decision problems

Suppose observations follow a distribution parametrised by $\theta \in \Omega$. Let $H : \theta \in \Omega'$ be the hypothesis that the effect lies in subset $\Omega'$ of $\Omega$. The hypothesis is associated with two decisions: the first, call it $d$, is that the parameter is not in $\Omega'$; the alternate decision is that the parameter is unknown. Choose a test statistic and a test alpha level. Then the procedure for deciding $d$ corresponds to whether the test statistic lies in the rejection region of a space $R$.

Lehmann shows that under certain conditions the decision defined by an intersection of negated hypotheses corresponds to the intersection of the rejection regions of the hypotheses. Formally, suppose we have a family of hypotheses $H_k : \theta \in (\Omega - \Omega_k)$ where $k$ belongs to a finite index set $I$. Let $R_k(\alpha_k) \subset R$ be the rejection region for a test of $H_k$ at level $\alpha_k$. For a given non-null subset $J$ of the index set, let $d_J$ be the decision $\theta \in \Omega_J$ where $\Omega_J = \cap_{k \in J} \Omega_k$. Call a decision *permissible* if $\Omega_J \neq \Omega_{J'}$ for any subset $J'$ of $I$ that contains $J$. Then the index set of a permissible decision must be of the form $J = \cup_{j \in J} \{k : \Omega_j \subseteq \Omega_k\}$. We will call such index sets *permissible*, defined as follows:

Given a family of hypotheses indexed by $I$ and a subset $J$ of $I$, $J$ is permissible if:

(a) $\cap_{k \in J} R_k (\alpha_k)$ is not empty, and

(b) for any $j, k \in I$ with $\Omega_j \subset \Omega_k$, then $j \in J \Rightarrow k \in J$.

Lehmann (1957b; p548) shows that for permissible sets $J$ there is a 1:1 correspondence between the rejection regions $D_J = \cap_J R_k (\alpha_k)$ and decisions $d_J$ that the true parameter $\theta \in \Omega_J = \cap_{k \in J} \Omega_k$. It is not necessary that the test set be coherent in the sense of requiring that whenever $H_j$ is rejected and $\Omega_j \subset \Omega_k$ then $H_k$ must also be rejected. Following Finner and Strassburger (2002; Theorem 2.2), coherence can be achieved by replacing the test of each $H_j$ with the maximum of the tests of hypotheses $H_k$ for which $\Omega_j \subseteq \Omega_k$. In our terminology, the new test of $H_j$ has rejection region $\cap_{k:\Omega_j \subseteq \Omega_k} R_k (\alpha_k)$ and by Berger (1982) it has level $\max \{ \alpha_k : \Omega_j \subseteq \Omega_k \}$. Under the new coherent set of tests, the decision $d_J$ that $\theta \in \cap_{k \in J} \Omega_k$ for some permissible subset $J$ corresponds to the intersection of the rejection regions for the new tests, which is

$$D_J = \cap_{j \in J} \left[ \cap_{k:\Omega_j \subseteq \Omega_k} R_k (\alpha_k) \right] = \cap_{j \in J} R_j (\alpha_j).$$

Here, the second equality follows from part (b) of the definition of permissible sets. Thus, the correspondence between rejection regions and permissible decisions applies regardless of whether the initial test set is coherent.

### 4.2. K-decision problems associated with sequences of effects

Lehmann (1957b; 10.1) considers families of hypotheses $H_k^- : \theta < \theta_k$ and $H_k^+ : \theta > \theta_k$ where $k \in \{1, ..., K\}$ and the parameter space is partitioned via a sequence

$$-\infty < \theta_1 < ... < \theta_K < \infty. \tag{1}$$

Such a family defines decisions of the form $\theta < \theta_1, \theta > \theta_K$ or $\theta_i < \theta < \theta_j$ for $i < j \in \{1, ..., K\}$. A 3-decision problem refers to the case $K = 1$ when the sequence (1) has only one finite value $\theta_1$. MBD concerns the 6-decision problem in which $K = 2$, as described by Lehmann. This has four generating hypotheses, which we denote by $H_k^+ : \theta > \theta_k$ and $H_k^- : \theta < \theta_k$ for $k = 1$ and 2.

The six decisions correspond to the six regions in Fig. 1b and are listed in the first column of Table 1. Alongside each decision are the hypotheses that must be rejected to determine it. We also list labels for these decisions used in Fig. 1 and the decision labels used in MBD (Hopkins and Batterham, 2016). Note that MBD documentation refers only to the hypotheses we denote as $H_1^-$ and $H_2^+$. Fig. 4 is another illustration of rejection regions in the 6-decision problem, this time when tests of $H_1^-, H_1^+$ are at level 0.005 and tests of $H_2^-, H_2^+$ are at level 0.25. This asymmetric testing creates a rejection region, shown striped, in which $H_2^-$ is rejected but $H_1^-$ is not. This is not a premissible decision, so under the theory it corresponds to decision $d_6$: inconclusive.

**Table 1.** Permissible decisions in the 6-decision problem and the hypotheses that must be rejected to determine them.

| Decision | Rejected tests | Fig. 1 terminology | MBD terminology |
|---|---|---|---|
| $d_1 : \theta < \theta_1$ | $H_1^+, H_2^+$ | Inferior | Negative |
| $d_2 : \theta < \theta_2$ | $H_2^+$ | Non-superior | Possibly negative |
| $d_3 : \theta_1 \leq \theta \leq \theta_2$ | $H_1^-, H_2^+$ | Equivalent to 0 | Trivial* |
| $d_4 : \theta > \theta_1$ | $H_1^-$ | Non-inferior | Possibly positive |
| $d_5 : \theta > \theta_2$ | $H_2^-, H_1^-$ | Superior | Positive |
| $d_6 : \theta \in \Omega$ | | Inconclusive | Unclear |

\* The MBD spreadsheets currently compare an alpha level with the sum of the *p*-values of the two test statistics rather than the maximum of these values. Earlier versions of the spreadsheets also subdivided decisions $d_2$ and $d_4$.

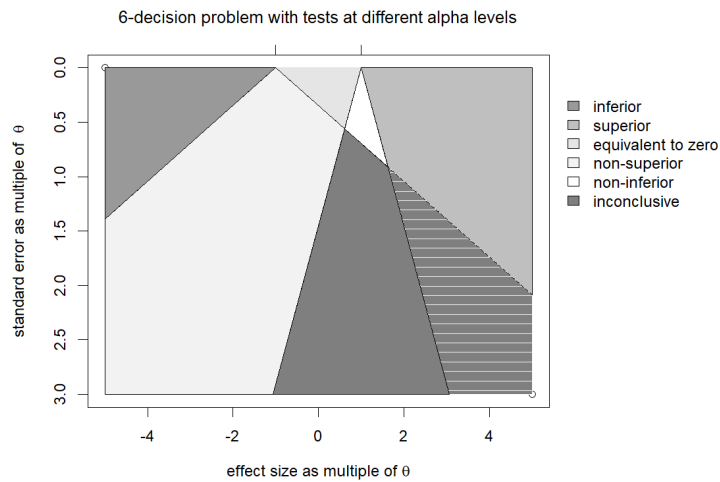

**Fig. 4.** Striped region does not correspond to a permissible decision.

### 4.3. A construction that extends Lehmann's theorem to test at multiple alpha levels, and its application to generalised funnel plots and MBD

To extend Lehmann's framework to allow testing of each hypothesis at multiple levels, first assume that testing is at two levels. Take a biased coin. From $H_2^+$ form two hypotheses, the first of which simply says that $\theta > \theta_2$ while the second says that $\theta > \theta_2$

OR the coin always comes up tails. Extend each of the other hypotheses likewise. Amend each observation by flipping the coin and adjoining the result as 0 or 1. Apply the standard test for the Bernoulli parameter, so that rejection regions are subsets in $R \times [0,1]$. This construction allows us to apply Lehmann's correspondence to the enlarged set of permissible decisions. Finally, use induction to extend the theory to tests at an arbitrary number of alpha levels. Mathematical details are in the Supporting Material.

The "6-decision" problem blows out to 27 permissible decisions when testing is at three alpha levels. Their corresponding rejection regions, projected onto $R$, are illustrated in Fig. 5 as the polygons formed by the linear boundaries of the various hypothesis rejection regions. As with the earlier figures, this figure assumes a $t$-test with 18 degrees of freedom and alpha levels of 0.25, 0.05 and 0.005. Boundaries of rejection regions of hypotheses $H_1^-$ and $H_1^+$ are shown as dotted lines and of $H_2^-$ and $H_2^+$ as solid black lines, with stronger lines and shades corresponding to tests at smaller alpha levels. Regions of the same shade group the rejection regions of tests at the same nominal alpha level. These grouped regions are precisely those of the generalised funnel plot depicted in Fig. 3a.

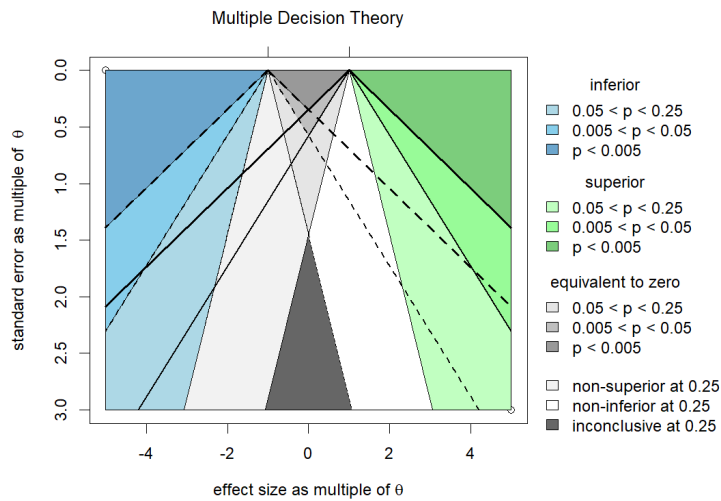

**Fig. 5**. Polygonal rejection regions for the 27 permissible decisions in the "6-decision" problem with testing at 3 levels; the colours group decisions with the same nominal alpha level.

As noted in the previous Section, MBD additionally constrains conclusive decisions to satisfy non-inferiority and non-superiority conditions. The Supporting Material spells out what hypotheses must be rejected for each MBD decision.

The spreadsheets optionally allow a nominated ratio of the $p$-values for benefit and harm to override an 'unclear' finding in clinical MBI. Use of this option is outside the theory. The current MBD decisions also differ from our description in the test for equivalence. As noted in Table 1, the spreadsheets determine the probabilistic term attached to a trivial decision using the sum of the *p*-values for testing non-inferiority and testing non-superiority. In our theory the maximum of these *p*-values determines whether an equivalence test is satisfied. Because the maximum is a more powerful test, where the spreadsheets differ from the theory they return a weaker finding, in the sense of determining 'likely trivial', say, rather than 'very likely trivial'. Note that the MBD formalisation by Welsh and Knight (2014) represented data points as triples (the two *p*-values and 1 minus their sum) that were plotted onto a ternary space to determine the decision. This modelling strips off the test levels, to leave only four decisions, and is not helpful for seeing where the computations sit theoretically.

## 5. Discussion and Conclusion

Greenland (2017) observes that controversies arise when a promoter of a new methodology fails to recognise its limitations and errors and at the same time critics misunderstand and misrepresent the proposal. Our positioning of MBD within the theoretical framework of multiple decision problems is the first full mathematical explanation of the procedure. This allowed us to position MBD as an extension of the 6-decision problem, in which decisions are the intersections of one-sided hypotheses that relate to a pair of values in a parameter space. MBD decisions are the intersections of similar one-sided hypotheses, although these hypotheses concern values on an extended parameter space in which the extensions effectively label the level of test applied to the hypothesis. We preserved the link between decisions and rejection regions (Lehmann, 1957a,b) that enables decisions to be made by plotting summary data onto a chart of rejection regions.

A tool to create such charts at nominated test levels, to enter data and to nominate context-appropriate labels is available at https://ja090.shinyapps.io/MBDSS. The tool allows Type I error rates to be estimated. It also serves to create generalised contour-enhanced funnel plots, which we showed are closely related to MBD. Visualisation tools have long been recognised as a mechanism to involve stakeholders in the research process and we suggest they also have a role in moving researchers away from dichotomous decisions based on statistical significance (Amrhein et al., 2019).

Findings for which the evidence is statistically weak still inform a decision maker of which of the possible decisions is most compatible with the data, and they serve the broader purpose of adding to a body of knowledge through their contribution to meta-analyses (Lakens, 2020). Arguments about misuse of Bayesian terminology or imprecision of MBD's verbal labels need to be evaluated against the goal of effective communication with stakeholders who are not statistically trained and for whom academic/scientific language may present a barrier. Attaching labels to ranges and categories is fundamental to human learning and reasoning. The most useful terminology for describing MBD decisions will depend on the context.

Our presentation does not match current MBD documentation, which generally portrays the procedure as semi-Bayesian or reference-Bayes without offering a mathematical formalisation. A Bayesian parallel to our approach would be worthwhile.

We also have not addressed limitations and errors within MBD documentation and usage that have been the subject of earlier criticisms. Additionally, MBD analyses have almost always used defaults to set parameters rather than exploit expert knowledge as we have proposed.

While work remains to convert MBD into a robust decision support tool that will engage stakeholders, a recent recommendation to simply replace MBD with one-sided tests (Aisbett et al., 2020) takes a Neyman-Pearson perspective that only captures part of MBD's potential application. Unqualified conclusions that the procedure "should not be used" (Welsh and Knight, 2014: p883; Sainani, 2018: p2166; Sainani et al 2019; p17) also reflect a narrow view. As Greenland (2017: p3) suggests, it is unwise to "reject out of hand any methodology because it is flawed or limited, for a methodology may perform adequately for some purposes despite its flaws." Our mathematically sound version of MBD, with its visualisation capabilities, has the potential to improve decision-making by helping scientists choose statistical analyses appropriate to their context and engage stakeholders in the interpretation of the data.

**SUPPORTING MATERIAL FOR 'ADVANCING STATISTICAL DECISION-MAKING IN SPORTS SCIENCE' BY AISBETT, DRINKWATER, QUARRIE & WOODCOCK (ADQW).**

## Extending the Multiple Decision Procedure to Test at Multiple Levels

We adapt Lehmann (1957a,b) to allow the same hypothesis to be tested at multiple alpha levels. Such testing corresponds to setting multiple loss functions for the same hypothesis, and thus corresponds to adopting different models of loss and risk. We describe the construction and establish that the correspondence between intersections of negated hypotheses and rejection areas can be maintained. Then we describe the MBD decisions in terms of rejected hypotheses.

Notation and assumptions are as in the main paper (ADQW).

### A1. Extending the parameter space and the family of hypotheses

Initially, suppose that the hypotheses in the family can be tested at two levels. To fit within Lehmann's framework as presented in ADQW: Section 2 we convert each hypothesis into two hypotheses using a nuisance parameter. Specifically, we extend the parameter space to $\Omega(1) = \Omega \times [0,1]$ and partition it into subsets $(\theta_{i-1}, \theta_i] \times [0,1], i = 1, K$ and $(\theta_K, \infty) \times [0,1]$. Next, for each $k \in I$ we adapt the hypothesis $H_k$ to assert that the parameter $(\theta, \pi) \in \Omega \times [0,1]$ satisfies $\theta \in \Omega - \Omega_k$ and makes no comment about the value of $\pi$, that is $(\theta, \pi) \in (\Omega - \Omega_k) \times [0,1]$. Denote this hypothesis by $H_k(0)$. We also form the hypothesis $H_k(1)$ which asserts that either $(\theta, \pi) \in (\Omega - \Omega_k) \times [0,1]$ or $\pi = 0$, i.e., $(\theta, \pi) \in ((\Omega - \Omega_k) \times [0,1]) \cup (\Omega \times \{0\})$. Obviously $H_k(0) \Rightarrow H_k(1)$ and the new family of hypotheses is indexed by the set

$$I(1) = \{(k,0), (k,1) : k \in I\}. \tag{A.1}$$

Rejecting $H_k(0)$ corresponds to $(\theta, \pi) \in \Omega_k \times [0,1]$. Rejecting $H_k(1)$ corresponds to $\theta \in \Omega_k$ and $\pi > 0$; that is, $(\theta, \pi) \in \Omega_k \times (0,1]$. Rejecting all hypotheses in a set $\{H_k(0), H_l(1) : k \in J, l \in J_1\}$ where $J$ and $J_1$ are subsets of $I$ therefore corresponds to the decision

$$d_{J, J_1} : (\theta, \pi) \in \cap_{k \in J} (\Omega_k \times [0,1]) \cap (\cap_{l \in J_1} (\Omega_l \times (0,1]))$$
$$= (\cap_{k \in J - J_1} \Omega_k \times [0,1]) \cap ((\cap_{l \in J_1} \Omega_l \times (0,1])). \tag{A.2}$$

(Here and elsewhere we use the convention that an intersection taken over an empty index set is the entire parameter space.)

*A2. Enlarging the sample space and defining rejection regions to apply Lehmann's Theorem 1*

The data are assumed to be $N$ observations of a random variable $X$ with parameter $\theta$. We inject the data into the product of the parameter space and the unit interval by taking each observation $x$ to the pair $(x, y_x)$ where $y_x$ is an observation of a Bernoulli random variable with parameter $\pi$, i.e., $y_x$ is 0 or 1 and $\sum_x y_x \sim B(N, \pi)$.

Let $R_k(\alpha_k) \subset R$ denote the rejection region of $H_k$ at test level $\alpha_k$. For each $k \in I$, label the two test levels so that $\alpha_k > \alpha_{k1}$ and hence $R_k(\alpha_{k1}) \subset R_k(\alpha_k)$. The hypothesis $: \pi \in [0,1]$ is always true, so $H_k(0)$ has rejection region $R_k(\alpha_{k1}) \times [0,1]$. The rejection region of the hypothesis $H : \pi = 0$ with test statistic $\sum_x y_x / N$ is $(0,1]$, for any test level $\alpha_k > 0$. Since $H_k(1)$ is the union of $H_k$ and $H$, it is rejected at level $\alpha_k$ whenever both these hypotheses are rejected at that level and its rejection region is the intersection of their rejection regions, namely $R_k(\alpha_k) \times (0,1]$ (Berger, 1982). Thus the rejection region of $H_k(1)$ is contained in the rejection region of $H_k(0)$, given the ordering of the test levels. By the definition of permissible subsets, a permissible subset of the index set $I(1)$ containing $(k,1)$ must therefore contain $(k,0)$. It follows that permissible subsets have the form $\{(k,0), k \in J\} \cup \{(l,0),(l,1), l \in J_1\}$ where $J$ and $J_1$ are permissible subsets or are null.

Given such subsets, from the above,

$$D_{J,J_1} = \left(\cap_{k \in J}\left(R_k(\alpha_k) \times [0,1]\right)\right) \cap \left(\cap_{k \in J_1}\left(R_k(\alpha_{k1}) \times (0,1]\right)\right)$$
$$= \left(\cap_{k \in J} R_k(\alpha_k) \times [0,1]\right) \cap \left(\cap_{k \in J_1} R_k(\alpha_{k1}) \times (0,1]\right). \tag{A.3}$$

We can now invoke Lehmann's correspondence applies between the rejection regions $D_{J,J_1}$ defined in (A3) and the decisions $d_{J,J_1}$ defined in ( A.2). There is thus a 1:1 correspondence

$$d_{J,J_1} \leftrightarrow \left(\cap_{k \in J} R_k(\alpha_k) \times [0,1]\right) \cap \left(\cap_{k \in J_1} R_k(\alpha_{k1}) \times (0,1]\right). \tag{A.4}$$

This correspondence would hold regardless of how the test alpha levels were ordered. However, the ordering means that $R_k(\alpha_{k1}) \subset R_k(\alpha_k)$ whenever $R_k(\alpha_{k1}) \times (0,1] \subset R_k(\alpha_k) \times [0,1]$, and so there is a further 1:1 correspondence

$$\left(\cap_{k \in J} R_k(\alpha_k) \times [0,1]\right) \cap \left(\cap_{k \in J_1} R_k(\alpha_{k1}) \times (0,1]\right) \leftrightarrow \left(\cap_{k \in J} R_k(\alpha_k)\right) \cap \left(\cap_{k \in J_1} R_k(\alpha_{k1})\right). \tag{A.5}$$

Thus the decision $d_{J,J_1}$ is made when the decision $d_J$ would have been made with testing at the less stringent level and also the decision $d_{J_1}$ would have been made with testing at the more stringent level.

There are now 14 permissible decisions for the "6-decision" problem: the 5 original permissible decisions, each at two possible levels, plus four decisions made with asymmetric test levels. These mixed-level tests decide $\theta_1 \leq \theta \leq \theta_2$ on rejecting hypotheses in the union $\left\{ H_1^-(0) \right\} \cup \left\{ H_2^+(0), H_2^+(1) \right\}$ or in the union $\left\{ H_2^+(0) \right\} \cup \left\{ H_1^-(0), H_1^-(1) \right\}$; decide $\theta < \theta_1$ on rejecting hypotheses in the union $\left\{ H_1^+(0), H_2^+(0) \right\} \cup \left\{ H_2^+(0), H_2^+(1) \right\}$; and decide $\theta > \theta_2$ on rejecting $\left\{ H_1^-(0), H_2^-(0) \right\} \cup \left\{ H_1^-(0), H_1^-(1) \right\}$. Note that these four "mixed-level" tests are at the highest alpha level of the component tests (Berger, 1982).

### A3. Extending the construction to further test levels

If testing is at three levels, we first extend the theory to testing at two levels using the construction described above, and then re-apply the construction to enlarge the parameter space $\Omega(1) = \Omega \times [0,1]$ to $\Omega(2) = \Omega \times [0,1] \times [0,1]$. Hypothesis $H_k$ tested at three levels is converted to the three hypotheses with negations

$$\neg H_k(0): (\theta, \pi_1, \pi_2) \in \Omega_k \times [0,1] \times [0,1],$$

$$\neg H_k(1): (\theta, \pi_1, \pi_2) \in \Omega_k \times [0,1] \times (0,1], \qquad \text{(A.6)}$$

$$\neg H_k(2): (\theta, \pi_1, \pi_2) \in \Omega_k \times (0,1] \times (0,1]$$

that are tested at decreasing alpha levels. By induction, we can handle hypotheses tested at $n$ levels. The index set for the enlarged family is $I(n) = \left\{ (k,m), k \in I, 0 \leq m \leq n \right\}$. We are associating $R$ with an expanded number of rejection regions, and hence an expanded number of decisions.

### A4. MBD decisions in terms of rejected hypotheses

The "6-decision" problem has 27 permissible decisions when testing is at three levels. MBD groups decisions concerning the same parameter range made at the same nominal level. Modifier verbal labels such as 'likely' indicate the test level of the superior/positive, inferior/negative and equivalent/trivial decisions; we have formally captured these labels in the values 0, 1 or 2 attached to the three hypotheses created against each of the original hypotheses. There are two exceptions to this grouping rule. Firstly, some decisions are lumped with 'inconclusive' into a decision called 'unclear', with the two variants of MBD differing in which decisions these are. Secondly, 'possibly positive' is decided when only the hypothesis $H_1^-$ is rejected at some level, and analogously for 'possibly negative'. The decisions and rejected hypotheses are listed in Table A1.

**Table A1.** 12 MBD decisions in terms of the "6-decision" problem with testing at 3 levels.

| Decision | Non-clinical MBD label | Sets of rejected hypotheses |
|---|---|---|
| $d_1(0): \theta < \theta_1$ | likely negative | $\{H_1^+(0), H_2^+(1)\}, \{H_1^+(0), H_2^+(2)\}$ |
| $d_1(1): \theta < \theta_1$ | very likely *negative* | $\{H_1^+(1), H_2^+(2)\}, \{H_1^+(1), H_2^+(1)\}$ |
| $d_1(2): \theta < \theta_1$ | most likely *negative* | $\{H_1^+(2), H_2^+(2)\}$ |
| $d_2: \theta < \theta_2$ | possibly negative (not pos've) | $\{H_2^+(1)\}, \{H_2^+(2)\}$ |
| $d_3(0): \theta_1 \le \theta \le \theta_2$ | likely *trivial* | $\{H_1^-(0), H_2^+(i)\}, \{H_1^-(i), H_2^+(0)\}, i=1,2$ |
| $d_3(1): \theta_1 \le \theta \le \theta_2$ | very likely *trivial* | $\{H_1^-(1), H_2^+(1)\}, \{H_1^-(1), H_2^+(2)\}, \{H_1^-(2), H_2^+(1)\}$ |
| $d_3(2): \theta_1 \le \theta \le \theta_2$ | most likely *trivial* | $\{H_1^-(2), H_2^+(2)\}$ |
| $d_4: \theta > \theta_1$ | possibly positive (not neg've) | $\{H_1^-(1)\}, \{H_1^-(2)\}$ |
| $d_5(0): \theta > \theta_2$ | likely positive | $\{H_1^-(1), H_2^-(0)\}, \{H_1^-(2), H_2^-(0)\}$ |
| $d_5(1): \theta > \theta_2$ | very likely *positive* | $\{H_1^-(2), H_2^-(1)\}, \{H_1^-(1), H_2^-(1)\}$ |
| $d_5(2): \theta > \theta_2$ | most likely *positive* | $\{H_1^-(2), H_2^-(2)\}$ |
| $d_6: -\infty < \theta < \infty$ | unclear | $\{H_2^+(0)\}, \{H_1^-(0)\}, \{H_1^-(0), H_2^-(0)\},$ $\{H_1^+(0), H_2^+(0)\}, \{H_1^-(0), H_2^+(0)\}, \varnothing$ |

In Table A1, $H_2^+(2)$ is $H_2^+$ tested at level 0.005, $H_1^-(0)$ is $H_1^-$ tested at level 0.25, and so on, assuming the MBD default alpha levels (Hopkins, 2007). A decision is made when all the rejected hypotheses lie in one of the sets in the corresponding entry in the 'Sets of rejected hypotheses' column. The sets each represent a permissible decision, although for conciseness a set containing $H_2^+(2)$, say, is not shown as containing $H_2^+(1)$ and $H_2^+(0)$ and these must be taken as implied.

Note that earlier versions of MBD had a decision called 'possibly trivial' not in Table A1. This decision was made according to $p$-values from the tests $H_1^-$ and $H_2^+$ and is not valid under the theory. What had been a 'possibly trivial' decision will be either 'possibly positive' or 'possibly negative' depending on whether $H_1^-(1)$ or $H_2^+(1)$ is rejected. The MBD implementations of the remaining 'trivial' decisions still needs to be amended to accord with the theory, as noted in ADQW: Table 1.

The decisions listed in Table A2 are annotated 'consider using' in clinical MBD; all other decisions are annotated 'don't use'. An 'unclear' decision is made for non-clinical MBD when neither $H_1^-(1)$ nor $H_2^+(1)$ is rejected and for clinical MBD when neither $H_1^-(2)$ nor $H_2^+(0)$ is rejected.

**Table A2**. Clinical MBD annotates some decisions as 'consider using'

| Decision | Clinical MBD label | Rejected hypotheses |
|---|---|---|
| $d_4 : \theta > \theta_1$ | possibly beneficial (not harmful) | $\left\{ H_1^-(2) \right\}$ |
| $d_5(0) : \theta > \theta_2$ | likely beneficial | $\left\{ H_1^-(2), H_2^-(0) \right\}$ |
| $d_5(1) : \theta > \theta_2$ | very likely *beneficial* | $\left\{ H_1^-(2), H_2^-(1) \right\}$ |
| $d_5(2) : \theta > \theta_2$ | most likely *beneficial* | $\left\{ H_1^-(2), H_2^-(2) \right\}$ |
| all other decisions | do not use | all other combinations |